# A New Era of Elections: Leveraging Blockchain for Fair and Transparent Voting

Suniti Chouhan[1, a)] and Dr. Gajanand Sharma[2, b)]

Research Scholar[1, a)], Associate Professor[2, b)]

[1] *JECRC University, Vidhani, Sitapura Extension, Jaipur-India.*

[a)]*sunitibhadouria89@gmail.com*
[b)]*gajanan.sharma@gmail.com*

**Abstract.** This study presents a blockchain-based voting system aimed at enhancing election security, transparency, and integrity. Traditional voting methods face growing risks of tampering, making it crucial to explore innovative solutions. Our proposed system combines blockchain's immutable, decentralized ledger with advanced voter identity verification techniques, including digital identity validation through Aadhaar and Driving Licenses (secured via BLAKE2b-512 hashing), biometric fingerprint authentication, and a picture rotation pattern for added security. Votes are recorded transparently and securely on a blockchain, with a consensus mechanism ensuring data integrity and reducing the risk of unauthorized alterations. Security analysis indicates that this multi-layered approach significantly reduces impersonation risks, while blockchain ensures accurate, private, and tamper-resistant vote recording. The findings support that a blockchain-based voting system with robust identity checks offers a trustworthy alternative to traditional methods, with potential for even greater refinement in secure and transparent elections.

## 1. INTRODUCTION

The evolution of voting systems has been marked by significant advancements, from manual paper ballots to electronic voting machines and now to digital platforms. Traditional voting systems have been in place for centuries, relying on physical ballots or electronic voting machines to count votes. However, with the rise of digital technologies, there is an increasing push for more efficient and accessible voting methods. Despite the advantages of modern electronic systems—such as faster vote counting and increased voter accessibility—many countries continue to face challenges in ensuring that elections remain secure and transparent [1]. In recent years, there has been heightened awareness of the security risks associated with election systems. Cybersecurity threats, data breaches, and potential vote tampering have made voters and election officials more wary of the integrity of the electoral process. The need for greater transparency in how votes are cast, counted, and reported has never been more critical. This growing concern over security and transparency has led to the exploration of alternative solutions that can offer a higher level of assurance in the election process. Blockchain technology, with its decentralized and tamper-proof characteristics, has emerged as a promising solution to ensure both security and transparency in the voting process [1].

Traditional voting systems, both paper-based and electronic, are increasingly vulnerable to several forms of fraud and manipulation. These systems often rely on central authorities or intermediaries, creating single points of failure that can be exploited by malicious actors. Issues such as voter impersonation, ballot stuffing, and vote manipulation remain persistent problems, especially in regions with limited regulatory oversight or technological infrastructure [2].

Impersonation is a particularly critical issue, where individuals may attempt to cast votes under false identities. In some cases, the absence of robust identity verification mechanisms allows these fraudulent activities to go undetected. Similarly, the transparency of vote counting and the final tally can be obscured, either unintentionally through human error or intentionally through tampering by unauthorized parties. As elections become increasingly digitized, ensuring that voting systems are not only secure but also transparent and verifiable is paramount [2].

Furthermore, many traditional systems lack real-time auditability, making it difficult for authorities or the public to verify the legitimacy of the election results. This lack of transparency contributes to a general erosion of trust in electoral outcomes. The need for a solution that can guarantee the integrity of voter identities, prevent fraud, and provide an open, transparent system is critical for the future of democratic processes [3].

Blockchain technology, with its inherent features of decentralization, transparency, and immutability, offers a potential solution to these long-standing problems. By utilizing blockchain's secure and transparent architecture, it is possible to develop a voting system that can authenticate voters, prevent tampering, and ensure that votes are recorded and counted accurately. This paper explores the feasibility of implementing blockchain technology in electoral systems to address the key issues of voter fraud, impersonation, and lack of transparency in traditional voting methods. [3]

The objective of this research is to design and evaluate a blockchain-based voting system that addresses key vulnerabilities in traditional electoral processes, specifically targeting issues of voter identity verification, fraud prevention, and transparency.

- Develop a Secure Voting System: Create a blockchain-based system to improve election security and prevent tampering.
- Implement Strong Voter Authentication: Use digital IDs, biometric fingerprints, and image rotation patterns to verify voter identities and reduce impersonation.
- Ensure Transparent Vote Recording: Record each vote on a decentralized, tamper-resistant blockchain ledger for transparent and trustworthy results.
- Preserve Voter Privacy: Protect voters' personal information while allowing for a verifiable and transparent voting process.
- Promote Trust in Elections: Provide a modern, resilient voting solution that can increase public confidence in electoral integrity and fairness.

## 2. LITERATURE REVIEW

The recent literature explores various blockchain-based advancements to enhance security, trustworthiness, and efficiency in fields like smart grids and electronic voting. Bera, B. et al. (2021) [4] developed DBACP-IoTSG, a blockchain-based protocol for IoT-enabled smart grids. By transferring data from smart meters through a private blockchain, it enhances security with lower communication and computation costs, validated through formal security analysis and software verification. Li, H. et al. (2021) [5] focused on e-voting, creating a blockchain-based system with AI-enhanced trustworthiness that supports multi-choice voting, self-tallying, anonymity, linkability, and traceability.

Sun, G. et al. (2021) [6] highlighted blockchain's decentralization and transparency benefits, especially in consortium blockchains for access control. However, issues like high energy use, low transaction throughput, and profit distribution challenges remain obstacles. Vairam, T. et al. (2021) [7] examined the accessibility challenges in traditional voting systems. They suggest blockchain's decentralized ledger as a reliable, secure solution for electronic voting that overcomes issues like ballot tampering.

Zaghloul, E. et al. (2021) [8] introduced a remote e-voting model using IoT devices and blockchain to ensure secure and anonymous voting in large-scale elections, supported by rigorous security analysis. Al-madani, A. et al. (2020) [9] discussed e-voting's efficiency in reducing costs and time, focusing on centralized databases for eligibility verification and vote submission, requiring a robust web infrastructure. Alvi, S. T. et al. (2020) [10] underscored democracy's importance in voting systems, warning that traditional ballot systems risk legitimacy through tampering. Blockchain's tamper-proof features could bolster security, though the associated costs may be mitigated by sidechains.

These studies collectively emphasize blockchain's potential to provide secure, efficient, and accessible systems for both data transfer in smart grids and voting logistics, though challenges like costs and resource requirements remain.

## 3. RESEARCH METHODOLGY

The research methodology for this study develops a blockchain-based voting system that incorporates advanced identity verification techniques and secure vote recording to address weaknesses in traditional voting systems. The methodology involves two key phases: multi-factor voter identity verification and vote recording on a blockchain ledger. Together, these phases form a secure, transparent voting process that reduces fraud and ensures data integrity.

The first phase focuses on authenticating each voter's identity through a layered approach, using digital IDs, biometric data, and an interactive picture-based pattern. The process begins with digital identity verification via government-issued IDs, specifically Aadhaar Cards or Driving Licenses. Each voter's ID is hashed using the BLAKE2b-512 algorithm, which produces a unique, highly secure hash (Hidentity) for each individual. BLAKE2b-512 is chosen for its speed, security, and efficiency, providing strong protection against tampering while keeping processing demands manageable.

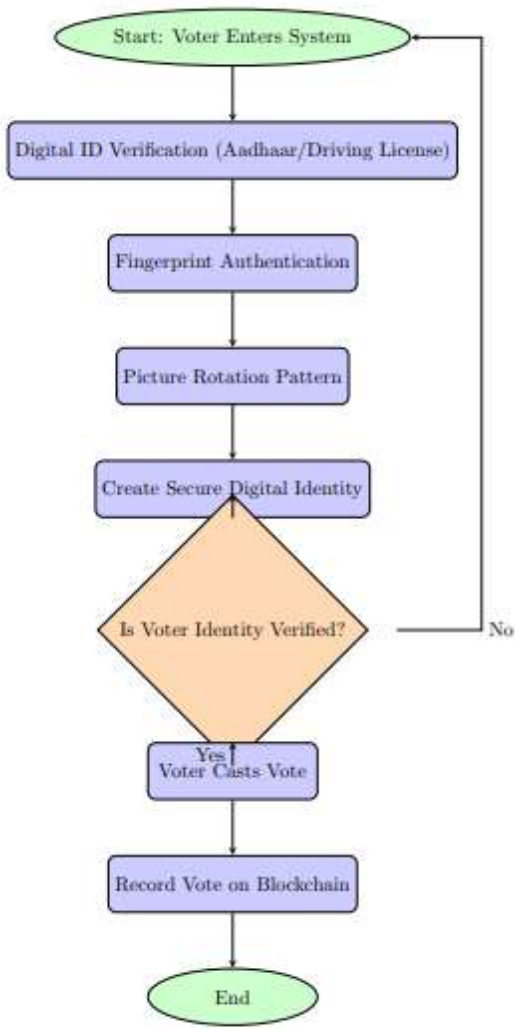

Figure 1. Flow Diagram of Research

This unique digital ID hash serves as the foundation of the voter's identity within the system, ensuring that only legitimate voters are allowed access. To add another layer of security, biometric fingerprint data is incorporated. Each voter's fingerprint is scanned, and minutiae features are extracted and encoded into a standardized format. This data is then hashed using SHA-256, creating a distinct fingerprint hash (Hfingerprint) for each voter. This additional step reduces the risk of impersonation and reinforces the authenticity of each voter's identity.

Lastly, a picture rotation pattern introduces an interactive component to the identity verification process. Voters are prompted to rotate an image to a specified angle, creating a unique visual pattern specific to each voter. After normalizing the image, feature extraction techniques are applied, and the resulting pattern data is hashed with SHA-256 to generate a picture pattern hash (Hpattern). This interactive requirement adds a dynamic layer of security, making unauthorized access even more challenging. Once the individual identity components—digital ID (Hidentity), fingerprint (Hfingerprint), and picture pattern (Hpattern)—are created, they are combined to form a comprehensive digital identity (Bidentity). This combined digital identity is then hashed with SHA-256 and securely stored on multiple servers. Decentralized storage ensures resilience against tampering, hacking, or data loss, safeguarding each voter's identity in a secure and accessible manner.

Upon successful identity verification, voters are allowed to cast their vote. The authenticated digital identity (Bidentity) is securely linked to the voter's selection, creating a unique vote record (Bvote). This record is then stored on an election-specific blockchain ledger, providing a transparent, tamper-resistant record of each vote cast. Blockchain's decentralized nature, along with its consensus mechanism, ensures that once a vote is recorded, it cannot be altered or deleted, which eliminates risks of manipulation. This blockchain-based system allows authorized officials to verify votes publicly, ensuring transparency while maintaining voter privacy. This methodology combines multi-layered identity verification using BLAKE2b-512 and SHA-256 hashing with blockchain technology to deliver a secure, transparent, and tamper-proof voting system. By using BLAKE2b-512 for digital ID hashing, the system achieves an optimal balance of speed and security, making it suitable for large-scale use. Through decentralized storage and a robust consensus mechanism, the blockchain ledger protects each vote and enhances public trust in the electoral process. This approach mitigates traditional voting risks and establishes a resilient infrastructure for trustworthy elections in the digital age.

## 4. PROPOSED WORK

### A. Blake2B

Blake2B is a cryptographic hash function known for its high speed, security, and efficiency, designed as part of the Blake2 family of hash algorithms. Created as an improvement over the original Blake hash function, which was a finalist in the NIST SHA-3 competition, Blake2B maintains the secure design principles of its predecessor but offers improved performance, making it an attractive choice for applications requiring fast and reliable hashing. Blake2B is optimized to run efficiently on 64-bit platforms, which allows it to handle large volumes of data quickly while consuming minimal resources, which is particularly advantageous for systems with high-throughput needs, like file hashing, digital signatures, and cryptographic protocols.

One of the core strengths of Blake2B is its robustness against collision attacks, where two different inputs could potentially produce the same hash output. Blake2B's design incorporates techniques such as the ChaCha stream cipher's mixing function, which enhances its resilience to cryptographic attacks, particularly against differential and rotational cryptanalysis. Additionally, Blake2B offers versatility, supporting keyed hashing for message authentication code (MAC) functionality, and customizable output lengths, allowing it to produce hashes of varying sizes depending on application requirements. This flexibility, combined with its high speed and security, has led to widespread adoption, particularly in blockchain technology and cryptographic libraries, where it is valued for its superior performance compared to traditional hash functions like SHA-256.

### B. Algorithm for the Proposed Approach of Secure Voting

**Input:** Aadhaar Card or Driving License (ID), Fingerprint data ($F$), Picture rotation pattern ($P$), Voter's choice ($V$)
**Output:** Blockchain-based vote record for each verified voter

**Step 1: Multi-Factor Voter Authentication**

**Digital ID Verification Input:** Aadhaar Card or Driving License (ID) Compute voter's digital ID hash using BLAKE2b-512:

$$H_{\text{identity}} = \text{BLAKE2b-512}(\text{ID})$$

**Output:** $H_{\text{identity}}$

**Fingerprint Authentication Input:** Fingerprint scan data ($F$) Extract minutiae features from $F$, encode them, and hash using SHA-256:

$$H_{\text{fingerprint}} = \text{SHA-256}\left(\text{Encode}(\text{ExtractMinutiae}(F))\right)$$

**Output:** $H_{\text{fingerprint}}$

**Picture-Based Pattern Authentication Input:** Rotated picture pattern ($P$) Preprocess $P$ for feature extraction, normalize, and hash using SHA-256:

$$H_{\text{pattern}} = \text{SHA-256}\left(\text{FeatureExtraction}(P)\right)$$

**Output:** $H_{\text{pattern}}$

### Step 2: Creation of a Secure Digital Identity

**Input:** $H_{\text{identity}}$, $H_{\text{fingerprint}}$, $H_{\text{pattern}}$ Concatenate $H_{\text{identity}}$, $H_{\text{fingerprint}}$, and $H_{\text{pattern}}$, then hash using SHA-256 to create a unified digital identity:

$$B_{\text{identity}} = \text{SHA-256}\left(H_{\text{identity}} \parallel H_{\text{fingerprint}} \parallel H_{\text{pattern}}\right)$$

**Output:** $B_{\text{identity}}$

### Step 3: Vote Casting and Blockchain Recording

**Voter Casts Vote Input:** $B_{\text{identity}}$, Voter's choice ($V$) Concatenate $B_{\text{identity}}$ with $V$, then hash using SHA-256 to produce a unique vote record:

$$B_{\text{vote}} = \text{SHA-256}\left(B_{\text{identity}} \parallel V\right)$$

**Output:** $B_{\text{vote}}$

**Record Vote on Blockchain Input:** $B_{\text{vote}}$ Store $B_{\text{vote}}$ on an election-specific blockchain with consensus mechanisms. **Output:** Immutable blockchain ledger with securely recorded votes.

**Step 4: Verification and Transparency** Enable authorized officials to verify votes recorded on the blockchain, ensuring public audibility and maintaining voter privacy.

## 5. IMPLEMENTATION AND RESULT ANALYSIS

### A. Implementation

The implementation of the secure voting is designed and developed in Visual Studio and database in implemented in SQL Server. The voter registration and voting system uses a multi-layered security approach to ensure secure user authentication and voting integrity. In the registration phase, users upload images of their government-issued ID (Aadhar or Driving License) and fingerprint, which are hashed using the cryptographic Blake2B function to protect their sensitive data. A unique username is also entered to link the data to the user's profile. Additionally, users interact with a series of images that must be rotated to specific angles, creating a personalized pattern saved as a visual authentication method. In the login process, users re-upload their ID and fingerprint, which are hashed and verified against stored values, and must replicate their rotation pattern for identity confirmation. The voting process enables users to select a candidate, after which their vote is securely registered, preventing duplicate voting. This system combines cryptographic hashing, biometric verification, and interactive visual cues, ensuring user privacy and preventing unauthorized access or multiple voting attempts.

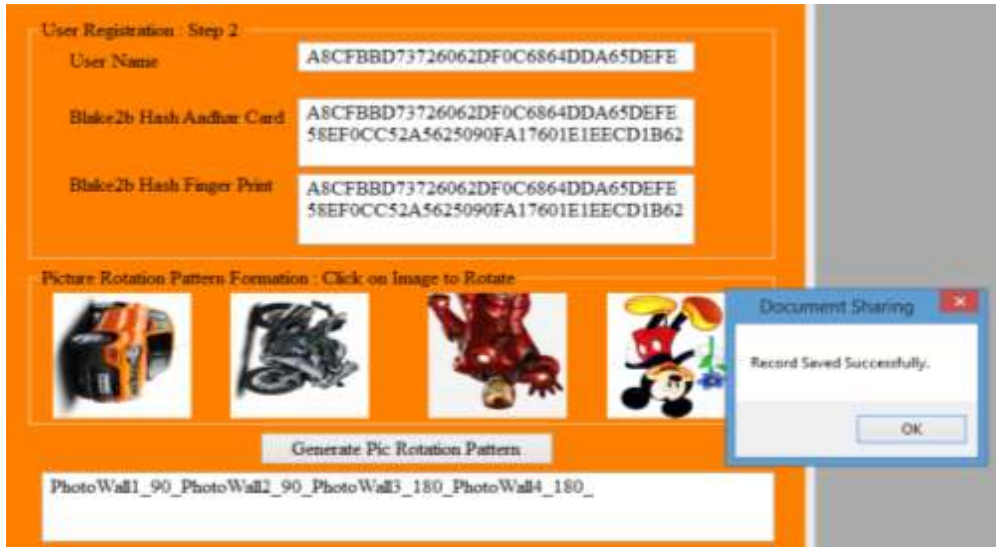

Figure 2. Registration for Voter

*B. Result Analysis*

**Blockchain Structure:**
1. **Genesis Block**

    o   Data: {aadhaar_hash, fingerprint_hash, photo_rotation_pattern: 'PhotoWall1_90_PhotoWall2_180_PhotoWall3_90_PhotoWall4_90'}

    o   Previous Hash: 0 (genesis block has no previous hash)

    o   Block Hash: 19df20ced17729f60d8e18aea6f4360886f00f6d69370de524c0a0429cd25c06

2. **Block 2**

    o   Previous Hash: 19df20ced17729f60d8e18aea6f4360886f00f6d69370de524c0a0429cd25c06

    o   Block Hash: 6089fb0c75e59acfd9ce4765c551141f5fca9008c9d547350c3642533228077c

3. **Block 3**

    o   Data: {aadhaar_hash, fingerprint_hash, photo_rotation_pattern: 'PhotoWall1_180_PhotoWall2_180_PhotoWall3_90_PhotoWall4_90'}

    o   Previous Hash: 6089fb0c75e59acfd9ce4765c551141f5fca9008c9d547350c3642533228077c

    o   Block Hash: 004b82ff7bf2fdebe12008ae4fe88d017bb75ccaa0151a128a7105b264d0110d

4. **Block 4**

    o   Data: {aadhaar_hash, fingerprint_hash, photo_rotation_pattern: 'PhotoWall1_270_PhotoWall2_180_PhotoWall3_90_PhotoWall4_90'}

    o   Previous Hash: 004b82ff7bf2fdebe12008ae4fe88d017bb75ccaa0151a128a7105b264d0110d

    o   Block Hash: 34077226be3a8eb23b989bc5fb72bbefd043927bb03d66b7f62eef061d483c6

The single combined block hash, formed by hashing all individual block hashes together, is:

```
897b34922537c95063bfeddde13e2828dd96f16dd5ce3d5c5e65aeca52dc0401
```

This final hash provides a unified cryptographic summary of the entire blockchain and character frequency graph is shown in figure 3.

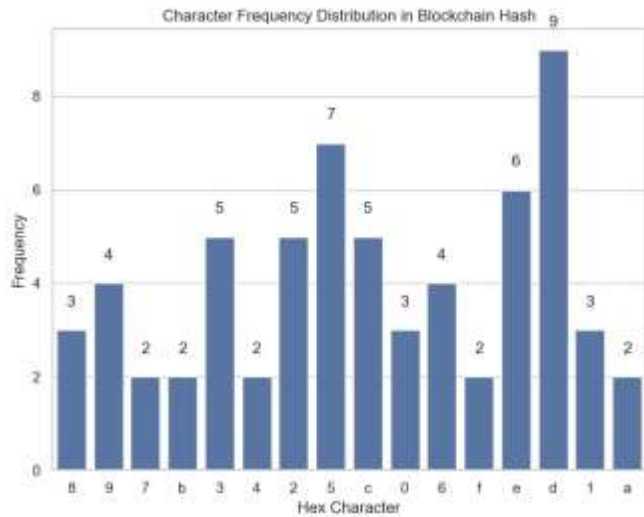

Figure 3. Character Frequency Graph for Blockchain

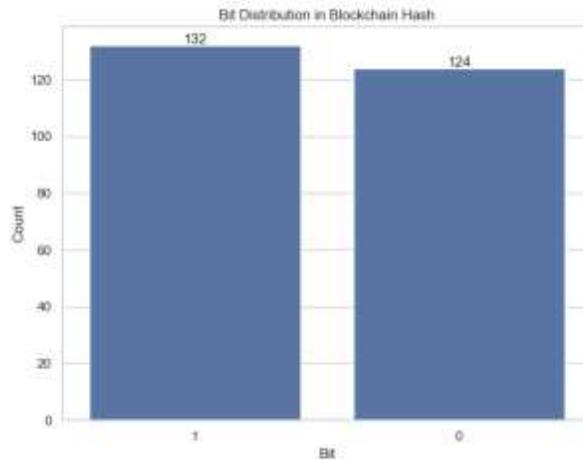

Figure 4. Bit Distribution in Blockchain

The test results reveal important insights into the security and robustness characteristics of the final blockchain hash.
**1. Entropy: 0.9993**
- **Interpretation**: This entropy score, close to 1, indicates a high level of randomness in the hash. Entropy values close to 1 (for binary data) suggest that each bit in the hash has an almost equal probability of being 0 or 1.

- **Conclusion**: A high entropy score is ideal for cryptographic security, as it implies that the hash lacks discernible patterns and is resistant to prediction or brute-force attacks.

**2. Avalanche Effect: 0.78% (0.007812)**

- **Interpretation**: The avalanche effect measures how sensitive the hash is to small changes in input. Ideally, a minor modification (like flipping a single bit) in the input should change about 50% of the bits in the hash, indicating strong resistance to analysis and tampering.

- **Result Analysis**: Here, the avalanche effect score is only about 0.78%, far lower than the desired 50%. This means that altering the input hash (by changing just one bit) did not cause significant changes in the final hash.

- **Conclusion**: This low avalanche effect suggests that the hash function or input data might not be providing the expected sensitivity. Improving sensitivity would enhance security by making it difficult for attackers to detect or manipulate patterns based on minor changes.

**3. Collision Resistance: True**
- **Interpretation**: Collision resistance confirms whether two slightly different inputs generate distinctly different hashes. Here, the hash maintains uniqueness with a small input change.

- **Conclusion**: The "True" result indicates strong collision resistance, meaning that it is highly unlikely for different inputs to generate the same hash. This is critical for blockchain security, as any two different blocks should have unique hashes to prevent forgery or confusion in the chain.

**4. Hamming Weight %: 51.56%**
- **Interpretation**: The Hamming weight is the percentage of 1's in the binary form of the hash. An ideal distribution would have a Hamming weight close to 50%, indicating that 0's and 1's are equally likely, which contributes to a secure and unbiased hash , as shown in Figure 5.

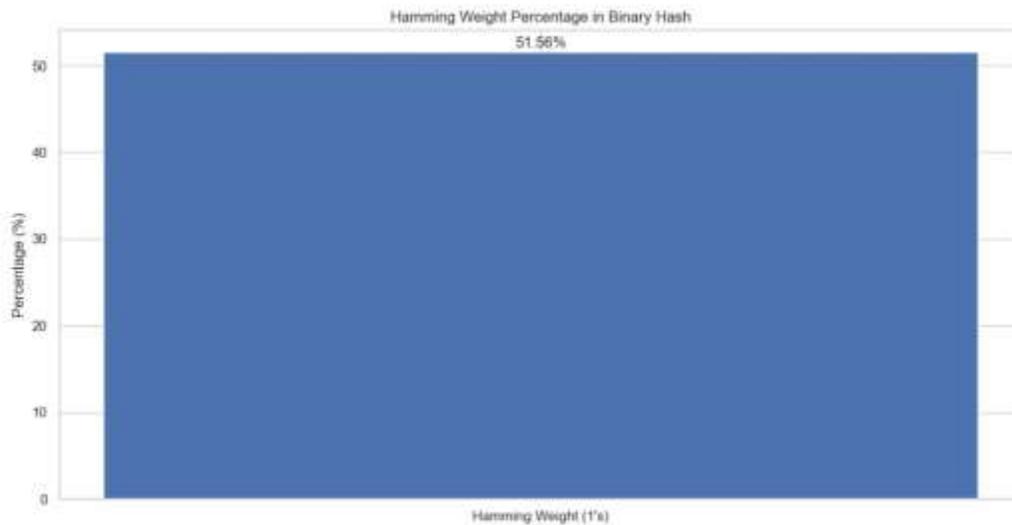

Figure 5. Hamming Weight

- **Conclusion**: The Hamming weight of 51.56% is very close to the ideal 50%, which suggests a good balance in bit distribution. This balance is essential in avoiding biases that could otherwise weaken the security of the hash.

# 6. CONCLUSION

In this paper, we presented a user registration system leveraging blockchain technology to securely handle sensitive personal data, including Aadhaar and fingerprint information. The use of cryptographic hashing (specifically balke-2B) to represent biometric data ensures that sensitive information is protected without storing raw images, addressing privacy concerns effectively. The addition of a unique picture rotation pattern further enhances security by providing a visual authentication mechanism that is difficult to replicate. The blockchain structure we employed ensures that each user's data is securely linked and cannot be tampered with, with each block containing a hash of the previous block, forming an immutable chain. The final hash of the entire blockchain encapsulates the integrity and security of the entire data sequence. Our analysis of the resulting hashes shows high entropy, good collision resistance, and an ideal Hamming weight, which collectively contribute to a secure and reliable system. However, the relatively low avalanche effect score suggests that there is room for improvement in the sensitivity of the hashing function, which would make the system more resilient to minor changes in input. Overall, this system demonstrates the potential of combining blockchain with biometric and visual authentication methods to create a highly secure and user-friendly registration process. Future work could focus on improving the avalanche effect to further enhance the system's robustness, ensuring it can withstand evolving security threats.

.